\pdfoutput=1
\documentclass[sigconf,nonacm]{acmart}

\AtBeginDocument{%
  \providecommand\BibTeX{{%
    \normalfont B\kern-0.5em{\scshape i\kern-0.25em b}\kern-0.8em\TeX}}}

\copyrightyear{2022}
\acmYear{2022}
\setcopyright{acmlicensed}\acmConference[AFT '22]{4th ACM Conference on Advances in Financial Technologies}{September 19--21, 2022}{Cambridge, MA, USA}
\acmBooktitle{4th ACM Conference on Advances in Financial Technologies (AFT '22), September 19--21, 2022, Cambridge, MA, USA}
\acmPrice{15.00}
\acmDOI{10.1145/3558535.3559784}
\acmISBN{978-1-4503-9861-9/22/09}

\usepackage{subcaption}
\usepackage{tikz}
\usepackage{balance}

\usepackage{tkz-kiviat} 
\usetikzlibrary{automata, positioning, arrows,calc,hobby,shapes,shapes.geometric,arrows,matrix}
\usetikzlibrary{shapes.multipart}
\usepackage{tkz-euclide}
\tikzstyle{int}=[draw, fill=lightgray, minimum height=1cm, minimum width=1.4cm,text width=1.3cm,align =center]
\definecolor{lightgreen}{RGB} {172, 243, 174}
\definecolor{lightred}{RGB}{250, 107, 132}
\definecolor{lightyellow}{RGB}{255, 255, 167}
\definecolor{lightgrey}{RGB}{220, 220, 220}
\tikzset{cross/.style={cross out, draw=black, ultra thick,fill=none, minimum size=4*(#1-\pgflinewidth), inner sep=0pt, outer sep=0pt}, cross/.default={3.5pt}}
\begin{document}

\title{SoK: Preventing Transaction Reordering Manipulations in Decentralized Finance}


\author{Lioba Heimbach}
\affiliation{%
\institution{ETH Zürich}
  \country{Switzerland}}
\email{hlioba@ethz.ch}
 
\author{Roger Wattenhofer}
\affiliation{%
\institution{ETH Zürich}
  \country{Switzerland}}
\email{wattenhofer@ethz.ch}


\begin{abstract}
User transactions on Ethereum's peer-to-peer network are at risk of being attacked. The smart contracts building \emph{decentralized finance (DeFi)} have introduced a new transaction ordering dependency to the Ethereum blockchain. As a result, attackers can profit from front- and back-running transactions. Multiple approaches to mitigate transaction reordering manipulations have surfaced recently. However, the success of individual approaches in mitigating such attacks and their impact on the entire blockchain remains largely unstudied. 

In this systematization of knowledge (SoK), we categorize and analyze state-of-the-art transaction reordering manipulation mitigation schemes. Instead of restricting our analysis to a scheme's success at preventing transaction reordering attacks, we evaluate its full impact on the blockchain. Therefore, we are able to provide a complete picture of the strengths and weaknesses of current mitigation schemes. We find that currently no scheme fully meets all the demands of the blockchain ecosystem. In fact, all approaches demonstrate unsatisfactory performance in at least one area relevant to the blockchain ecosystem.
\end{abstract}




\maketitle

\section{Introduction}
The introduction of Bitcoin~\cite{nakamoto2008bitcoin} sparked a wave in interest in cryptocurrencies and led to rapid development. Most notably Ethereum, which introduced \emph{smart contracts}~\cite{wood2014ethereum} and opened further opportunities for cryptocurrency development. Yet, cryptocurrencies only offered niche applications. The introduction of \emph{decentralized finance (DeFi)} on the Ethereum blockchain introduced a new purpose to cryptocurrencies. DeFi offers many traditional financial services without intermediaries, such as banks, brokerages, and stock exchanges. Instead of relying on intermediaries, DeFi utilizes smart contracts running on the blockchain.  

The smart contracts behind DeFi are often \emph{transaction order dependent}, i.e., the outcome of a set of transactions is dependent on their order. Therefore, DeFi gives rise to transaction reordering attacks -- race condition attacks. Consider the following example of a transaction reordering attack: you want to buy a cryptocurrency at a specific price; determined by the current state of the smart contract. A transaction reordering attack would change the cryptocurrencies price (for the worse) by interacting with the smart contract before your transaction executes, but after seeing your transaction. 

As simple financial transactions between two parties are not transaction order dependent, the traditional blockchain design does not pay much attention to transaction ordering. Instead, miners have complete control over transaction ordering. Users broadcast their transactions across the network. When building a block, miners not only choose the transactions included in a block and their order but can also include any additional transactions they wish. The freedom given to miners in building blocks gives rise to \emph{blockchain extractable value (BEV)} which defines the potential revenue from all kinds of transaction reordering attacks. More specifically, BEV is a measure of the profit that can be made through including, excluding, or re-ordering transactions within blocks~\cite{daian2020flash}. We note that BEV was previously known as \emph{miner extractable value (MEV)}.

Major DeFi applications, such as \emph{decentralized exchanges (DEXes)} and \emph{lending protocols} are susceptible to transaction reordering attacks. The monthly BEV collected on lending platforms and DEXes repeatedly exceeds \$100M~\cite{2022flashbotsmev} and presents an invisible tax to traders. This invisible tax on traders is further increased by the increased gas price resulting from BEV opportunities~\cite{daian2020flash}. Many \emph{BEV mitigation} schemes are currently under development;  they aim to alleviate traders from financial losses stemming from BEV. Their effects, however, must not be limited to BEV mitigation. Most schemes also intrude into other blockchain aspects, such as decentralization and goodput (the number of genuine transactions per block). Despite the beginning adaptation of BEV mitigation schemes, a full evaluation of their effects on the blockchain is missing.  

In this systematization of knowledge (SoK), we systematically categorize and evaluate state-of-the-art BEV mitigation schemes. In particular, we do not restrict our analysis to an approach's success in mitigating BEV but extend it to its full impact on the blockchain. Thereby, we aim to provide a balanced overview of the strengths and weaknesses of BEV mitigation schemes. Our work finds that no BEV mitigation scheme can meet all demands of the blockchain ecosystem as all schemes exhibit unsatisfactory performance in at least one area relevant to the blockchain ecosystem. Further research is, thus, required to avoid transaction reordering attacks without major negative disruptions to the blockchain ecosystem.

\section{Related Work}

Transaction reordering manipulations, in particular front-running, are not exclusive to blockchains. Front-running has long been prevalent in traditional finance. However, due to the non-public nature of traditional finance, front-running there is generally speculative~\cite{BERNHARDT2008front,angel2011equity}. Even more so, most forms of front-running are outlawed in traditional finance~\cite{markham1988front,moosa2015regulation}. Still, HFTs firms have gained worldwide attention for utilizing trading strategies that make use of what is sometimes considered \emph{legalized front-running}~\cite{harris2013what,scopino2014the}. While loopholes allowing forms of front-running prevail in traditional finance, lawmakers mitigate most front-running practices by outlawing the practice. Front-running can not be prevented through similar measures in DeFi and, therefore, DeFi requires new solutions. 

Front-running practices on DeFi were first studied by Eskandir et al.~\cite{eskandari2019sok}. They combine a scattered body of work surrounding DeFi front-running. Daiain et al.~\cite{daian2020flash} extend the research on DeFi transaction reordering manipulation by including all types of transaction reordering manipulations and introducing BEV. Note that at the time, BEV was known as miner extractable value. Further, they examine \textit{price gas auctions} (PGA) on the Ethereum blockchain. Qin et al.~\cite{qin2021quantifying} follow this line of research and are the first to quantify the transaction ordering tax -- showing that miners are actively extracting BEV. While these works identify BEV opportunities in DeFi, we systematize approaches that mitigate BEV. 

Baum et al.~\cite{baum2021mitigation} systematize and discuss front-running mitigation in DeFi. The work limits the discussion of the mitigation schemes to their ability to prevent front-running on DEXes. In contrast, our work is broader and systematizes BEV mitigation strategies. Further, we do not only assess a scheme's ability to prevent front-running but also consider the scheme's impact on the entire blockchain ecosystem. 

We also note that Baum et al.~\cite{baum2021mitigation} place a particular focus on a more general kind of front-running mitigation in their systematization. These more general front-running mitigation approaches do not focus on transaction reordering itself but develop new DEXes such that even after a transaction is confirmed, one cannot infer its size~\cite{chu2021manta,da2021kicking,li2021HoneyBadgerSwap,baum2021p2dex}. The goal, however, is mainly to provide confidentiality to users when trading. Confidentiality allows professional traders to execute their trading strategies unobserved -- preventing others from copying their strategies. While confidentiality could prevent speculative sandwich attacks, whereby the attacker guesses the transaction size based on the sender, we are not aware of such attacks taking place on DEXes. Further, these attacks are not a direct consequence of blockchain's public nature, as similar attacks are known from traditional finance~\cite{lewis2014flash}. We focus on prevention schemes for transaction reordering attacks extracting BEV on DeFi that do not impede the blockchain's public nature. 

Additionally, we note that protocols distributing the profits from transaction reordering attacks have recently emerged. Rook~\cite{2022rook} shares a proportion of the BEV collected from trades that execute through the protocol with the traders. Thus, while the tax placed on traders as a result of transaction reordering attacks is reduced, the practice is not irradiated. B.Protocol~\cite{2022bprotocol}, on the other hand,  pools the liquidity from individuals to work together and collect profits from liquidations. As these protocols generally target the distribution of BEV profits as opposed to mitigating BEV, we will not focus on them in the following.

\section{Transaction Reordering Manipulations}

We first introduce the basics of the Ethereum blockchain. Afterwards, we provide an overview of the kind of transaction reordering manipulation used on DEXes and lending protocols to extract BEV. 

Most DeFi applications run on the public Ethereum blockchain. To execute an Ethereum transaction, users first broadcast their transaction across Ethereum's peer-to-peer network. Thereby, the transaction enters the \emph{mempool} -- the public waiting area. Users indicate the \emph{gas fee} (Ethereum's network transaction fee) they are willing to pay when submitting a transaction. A transaction executes once a miner includes it in a block. Note that miners tend to prioritize transactions with higher gas fees, as they collect part of the fee. 

Due to the public nature of the Ethereum blockchain, any full Ethereum node can observe transactions in the mempool before they execute. Therefore, an attacker first sees an incoming transaction and then profits by inserting their transaction and manipulating the transaction ordering. We note that there are several strategies utilized by attackers to manipulate the transaction ordering to achieve their desired ordering. While the attacker can set the gas price to influence the transaction ordering, miners tend to order transactions according to the gas price, the attacker can also be the miner themselves or bribe the miner to order the transactions accordingly. 

\begin{figure}[ht]
\centering
\def\rectwidth{0.8}
    \begin{subfigure}{0.49\linewidth}
      \centering
        \begin{tikzpicture}[scale = 0.64]
        \begin{scope}[shift={(0,0)}]
            \draw[draw,thick, rounded corners = 2pt,fill= lightgrey] (-0.936,-0.936) rectangle ++(6.37,1.8) ;
            \node  (rect) at (0,0) [draw,fill= lightgreen,thick,minimum width=\rectwidth cm,minimum height=\rectwidth cm, rounded corners = 2pt] {\small $T_1$} ;
            \node  (rect) at (1.5,0) [draw,fill= lightgreen,thick,minimum width=\rectwidth cm,minimum height=\rectwidth cm, rounded corners = 2pt] {\small $T_V$};
            \node  (rect) at (3,0) [draw,fill= lightgreen,thick,minimum width=\rectwidth cm,minimum height=\rectwidth cm, rounded corners = 2pt] {\small $T_2$};
            \node  (rect) at (4.5,0) [draw,fill= lightgreen,thick,minimum width=\rectwidth cm,minimum height=\rectwidth cm, rounded corners = 2pt] {\small $T_3$};
        \end{scope}
    \end{tikzpicture}
    \caption{no attack} \label{fig:noattak}
  \end{subfigure}%
  \hfill
  \begin{subfigure}{0.49\linewidth}
  \centering
    \begin{tikzpicture}[scale = 0.64]
   
    \begin{scope}[shift={(7,0)}]
        \draw[draw, thick, rounded corners = 2pt,fill= lightgrey] (-0.936,-0.936) rectangle ++(6.37,1.8) ;
        \node  (rect) at (0,0) [draw,thick,fill= lightgreen,minimum width=\rectwidth cm,minimum height=\rectwidth cm, rounded corners = 2pt] {\small $T_1$};
        \node  (rect) at (1.5,0) [draw,thick,fill= lightgreen,minimum width=\rectwidth cm,minimum height=\rectwidth cm, rounded corners = 2pt] {\small $T_A$};
        \node  (rect) at (3,0) [draw,thick,fill= lightred,minimum width=\rectwidth cm,minimum height=\rectwidth cm, rounded corners = 2pt] {\small $T_V$};
        \draw (3.54,0.54) node[cross] {};
        \node  (rect) at (4.5,0) [draw,thick,fill= lightgreen,minimum width=\rectwidth cm,minimum height=\rectwidth cm, rounded corners = 2pt] {\small $T_2$};
    \end{scope}
    \end{tikzpicture}
    \caption{fatal front-running} \label{fig:fatalfrontrun}
  \end{subfigure}\vspace{0.4cm}

  \begin{subfigure}{0.49\linewidth}
  \centering
    \begin{tikzpicture}[scale = 0.64]
    \begin{scope}[shift={(7,0)}]
        \draw[draw, thick, rounded corners = 2pt,fill= lightgrey] (-0.936,-0.936) rectangle ++(6.37,1.8);
        \node  (rect) at (0,0) [draw,thick,fill= lightgreen,minimum width=\rectwidth cm,minimum height=\rectwidth cm, rounded corners = 2pt] {\small $T_1$};
        \node  (rect) at (1.5,0) [draw,thick,fill= lightgreen,minimum width=\rectwidth cm,minimum height=\rectwidth cm, rounded corners = 2pt] {\small $T_A$};
        \node  (rect) at (3,0) [draw,thick,fill= lightgreen,minimum width=\rectwidth cm,minimum height=\rectwidth cm, rounded corners = 2pt] {\small $T_V$};
        \node  (rect) at (4.5,0) [draw,thick,fill= lightgreen,minimum width=\rectwidth cm,minimum height=\rectwidth cm, rounded corners = 2pt] {\small $T_2$};
    \end{scope}
    \end{tikzpicture}
    \caption{front-running} \label{fig:frontrun}
  \end{subfigure}%
  \hfill
  \begin{subfigure}{0.49\linewidth}
  \centering
    \begin{tikzpicture}[scale = 0.64]
    \begin{scope}[shift={(7,0)}]
        \draw[draw, thick, rounded corners = 2pt,fill= lightgrey] (-0.936,-0.936) rectangle ++(6.37,1.8);
        \node  (rect) at (0,0) [draw,thick,fill= lightgreen,minimum width=\rectwidth cm,minimum height=\rectwidth cm, rounded corners = 2pt] {\small $T_1$};
        \node  (rect) at (1.5,0) [draw,thick,fill= lightgreen,minimum width=\rectwidth cm,minimum height=\rectwidth cm, rounded corners = 2pt] {\small $T_V$};
        \node  (rect) at (3,0) [draw,thick,fill= lightgreen,minimum width=\rectwidth cm,minimum height=\rectwidth cm, rounded corners = 2pt] {\small $T_A$};
        \node  (rect) at (4.5,0) [draw,thick,fill= lightgreen,minimum width=\rectwidth cm,minimum height=\rectwidth cm, rounded corners = 2pt] {\small $T_2$};
    \end{scope}
    \end{tikzpicture}
    \caption{back-running} \label{fig:backrun}
  \end{subfigure}%
\caption{A visualization of transaction order manipulation strategies. Figure~\ref{fig:noattak} shows a block with no attack, while Figures~\ref{fig:fatalfrontrun},~\ref{fig:frontrun}, and~\ref{fig:backrun} show the various reordering manipulations. We color a transaction that executed successfully green and a transactions that fail to execute are colored red and marked with a cross.} \label{fig:attacks}\vspace{0pt}
\end{figure}
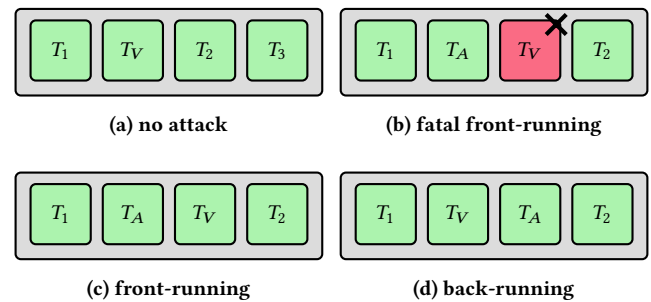

We outline the three kinds of transaction reordering manipulations (cf. Figure~\ref{fig:attacks}) required by the most common BEV opportunities:
\paragraph{Fatal front-running} Fatal front-running (cf. Figure~\ref{fig:fatalfrontrun}) describes a transaction reordering manipulation by which the attacker's transaction $T_A$ front-runs (executes before) the victims transactions $T_V$. In the process, the attacker's transaction causes the victim's transaction to fail.
\paragraph{Front-running} Front-running (cf. Figure~\ref{fig:frontrun}) is a transaction reordering manipulation that has attacker's transaction $T_A$ front-run the victims transactions $T_V$. As opposed to fatal front-running, the attacker ensures that the victim transaction will still execute. Generally, the conditions for the victim transaction, will however, be worse. 
\paragraph{Back-running} Back-running  (cf. Figure~\ref{fig:backrun}) occurs when the attacker's transaction $T_A$ back-runs (executing after) the victims transactions $T_V$. \\

Note that there is a fourth type of transaction reordering manipulation that is sometimes considered: \emph{clogging}. When clogging, the attacker fills up the block(s) until a deadline is reached to prevent others from executing a trade before the deadline. We, however, view this practice as a specific case of fatal front-running where the attacker front-runs the victim with multiple transactions.

In addition, to potentially worsening the conditions for the victim transaction, transaction reordering manipulations also increase the gas fee paid by the other transactions. Transaction reordering manipulations compete with other transactions for space in the block, thereby increasing the gas fee for other transactions. Further, the associated BEV opportunities are also known to cause PGAs, whereby multiple attackers compete for a BEV opportunity by offering higher and higher gas prices. The overbidding, in turn, increases the gas fee for the network's other transactions. 

We will continue by outlining the general mechanism of DEXes and lending protocols and the common BEV opportunities they present. We further note which transaction order manipulations these attacks require.

\subsection{Automated Market Makers}

Traditionally, centralized exchanges utilize the limit order book mechanism to facilitate trades. The limit order book mechanism matches individual buyers and sellers. \emph{Decentralized exchanges (DEXes)}, on the other hand, most commonly implement an \emph{automated market maker (AMM)}~\cite{hertzog2017bancor}. The four biggest DEXes on the Ethereum blockchain: Curve~\cite{2021curve}, Uniswap~\cite{2021uniswap}, Sushiswap~\cite{2021sushiswap}, and Balancer~\cite{2021balancer}, are all AMMs. More specifically, they function as \emph{constant function market makers (CFMMs)}~\cite{angeris2020improved}. Similar to their centralized counterparts, CFMMs allow users to exchange cryptocurrencies with each other. However, instead of utilizing the limit order book mechanism, CFMMs facilitate automatic algorithmic trading of cryptocurrencies. The CFMM only ensures that its predefined function stays constant. The simplicity of the CFMM trading mechanism is also the cause of their popularity on the Ethereum blockchain, as CFMM trades require little space on-chain. Thus, they utilize less gas and have relatively low transaction fees. While CFMMs have established themselves, there are transaction reordering attacks specific to them. To illustrate how transaction reordering manipulations on CFMMs can generate BEV, we focus on the most widely adopted subclass of CFMMs: \emph{Constant Product Market Makers (CPMMs)} in the following. 

A CPMM aggregates liquidity for every tradeable cryptocurrency pair in smart contracts -- known as \emph{liquidity pools}. Anyone can choose to provide liquidity in such a liquidity pool and, thereby, become a liquidity provider by depositing the pool's two assets (at equal value) in the liquidity pool. The pool's liquidity then facilitates trading between the two cryptocurrencies, and liquidity providers receive a small fee for every trade that utilizes their liquidity. The exchange rate offered to a transaction is algorithmically determined and ensures that the product of the amounts of the pool's reserves stays constant. Besides the trade size, the pool's size and the ratio between the pool's tokens set the exchange rate. In a CPMM, a trader wishing to exchange $\delta_x$ tokens $X$ for tokens $Y$ in the $X\rightleftharpoons  Y$ liquidity pool with reserves $x_t$ tokens $X$ and $y_t$ tokens $Y$ at time $t$, receives
\begin{equation*}
        \delta  _{y}= y_{t} - \frac{x_{t} \cdot y_{t}}{x_{t}+(1-f)\delta_x} = \frac{y_{t} (1-f )\delta_x}{x_{t}+(1-f)\delta_x}, 
\end{equation*}
tokens $Y$ \cite{adams2020uniswap}. In the previous, $f$ is the pool fee charged by the pool's liquidity providers. The collected transaction fee is distributed pro-rata to the liquidity providers. We note that the price per token $Y$, which is given by
$$\frac{x_t +(1-f) \delta _x}{y (1-f)},$$
increases with the trade input size $\delta_x$. As a result of the convexity of the price curve, the price per desired token increases with the trade size. 

\begin{figure}[t]
      \centering  
    \begin{subfigure}[]{0.48\linewidth}
    \centering  
    \definecolor{color1}{HTML}{8EB1C7}
\definecolor{color2}{HTML}{FFBF69}
\definecolor{color3}{HTML}{B02E0C}
\definecolor{color4}{HTML}{B744B8}
\definecolor{color5}{HTML}{610345}
\definecolor{color6}{HTML}{044B7F}
\definecolor{color7}{HTML}{8E6C88}

\begin{tikzpicture}[scale=1.6]

  \draw[-latex, thick] (0, 0) -- (2, 0) node[midway, below,sloped] {\small tokens $X$};
  \draw[-latex, thick] (0, 0) -- (0, 3.8) node[midway, above,sloped] {\small tokens $Y$};
  \draw[scale=1,line width=0.3mm, domain=0.555:1.9, smooth, variable=\x, gray] plot ({\x}, {2/\x});
  
   \node at (0.71,2.8)[circle,fill,inner sep=1pt](A) {} ;
   \node at (1.41,1.41)[circle,fill,inner sep=1pt](C) {} ;
   
   \draw [-latex,  line width=0.3mm, black] (A) to [bend right]  node[midway,left, inner sep = 7pt,text=black] {\small $T_V$}(C) ;

\end{tikzpicture}\vspace{0pt}
    \caption{execution of trade $T_V$ without a sandwich attack} \label{fig:sandwich0}
    \end{subfigure}
    \hfill
    \begin{subfigure}[]{0.48\linewidth}
    \centering  
    \usetikzlibrary{arrows}
\thispagestyle{empty}
\makeatletter
\pgfkeys{/kiviat/label style/.style={align=left,anchor=180+360/\tkz@kiv@radial*\rang}}

\usetikzlibrary{decorations.pathreplacing, arrows, fit}

\definecolor{color1}{HTML}{8EB1C7}
\definecolor{color2}{HTML}{FFBF69}
\definecolor{color3}{HTML}{B02E0C}
\definecolor{color4}{HTML}{B744B8}
\definecolor{color5}{HTML}{610345}
\definecolor{color6}{HTML}{044B7F}
\definecolor{color7}{HTML}{8E6C88}

\begin{tikzpicture}[scale=1.6]

  \draw[-latex, thick] (0, 0) -- (2, 0) node[midway, below,sloped] {\small tokens $X$};
  \draw[-latex, thick] (0, 0) -- (0, 3.8) node[midway, above,sloped] {\small tokens $Y$};
  \draw[scale=1,line width=0.3mm, domain=0.555:1.9, smooth, variable=\x, gray] plot ({\x}, {2/\x});
   \node at (0.71,2.8)[circle,fill,inner sep=1pt](A) {} ;
   \node at (0.896,2.219)[circle,fill,inner sep=1pt](C) {} ;
   
   \node at (1.594,1.247)[circle,fill,inner sep=1pt](B) {} ;
   \node at (1.088,1.827)[circle,fill,inner sep=1pt](D) {} ;
   \draw [-latex,line width=0.3mm,lightred] (A) to [bend right] node[midway,left,text = black] {\small$T_F$} (C) ;
   \draw [-latex,line width=0.3mm,black] (C) to [bend right] node[pos =0.58,left,inner sep=6pt,text = black] {\small$T_V$}(B) ;
   \draw [-latex, line width=0.3mm,lightred] (B) to [bend right] node[pos = 0.4,right,inner sep=6pt,text = black] {\small$T_B$}(D) ;

\end{tikzpicture}\vspace{0pt}
    \caption{execution of trade $T_V$ with a sandwich attack} \label{fig:sandwich1}
    \end{subfigure}
    \caption{Sandwich attack visualisation in a CPMM. Transaction $T_V$ is the victim's transaction, and transactions $T_F$ and $T_B$ are the attackers front- and back-running transactions respectively. Notice that the transaction $T_V$ receives a worse price in the presence of the sandwich attack.}\label{fig:sandwich}\vspace{0pt}
\end{figure}
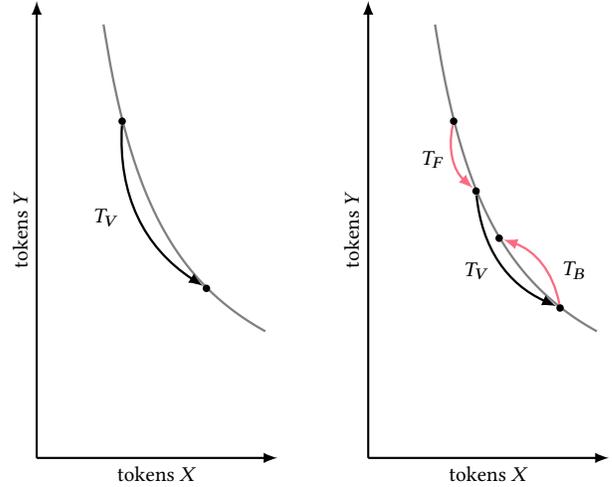

A CPMM transaction first broadcasts through the peer-to-peer network and enters the Ethereum mempool. The transaction, however, only executes upon inclusion in a block by a miner. Thus, the liquidity pool's state can change between the submission and execution time, thereby altering the exchange ratio received by the trade. Specifically, the pool's reserves $x_t$ and $y_t$ change with every transaction executing in the meantime. Thus, traders specify their \emph{slippage tolerance} -- the maximum price movement they are willing to accept. In case the price movement exceeds the specified tolerance, trade execution fails. Note that a high slippage tolerance puts the trader at risk of her transaction executing a worse price, while a low slippage tolerance puts her at risk of an unnecessary transaction failure. Nonetheless, the time between a trade's submission and execution makes it susceptible to transaction reordering attacks.

\emph{Sandwich attacks} are a common transaction reordering attack on CPMMs~\cite{zhou2021high}. A sandwich attack front-runs (cf. Figure~\ref{fig:frontrun}) and back-runs (cf. Figure~\ref{fig:backrun}) a CPMM transaction. Consider a victim's transaction wishing to trade 50 tokens $X$ for tokens $Y$ in transaction $T_V$. The pool fee is 0.3\%, and the transaction is submitted at time $t$. The pool reserves at time $t$ are 50 tokens $X$ and 200 tokens $Y$. If the transaction executes without the pool state changing, the trader receives 99.849 tokens $Y$. We visualize the execution of the transaction along the price curve in Figure~\ref{fig:sandwich0}. However, an attacker might observe the transaction in the mempool, waiting to be included in a block, and determine that it is profitable to perform a sandwich attack on the victim's trade. The attacker then front-runs the victim's trade with transaction $T_F$ buying 68.286 tokens $Y$ to inflate $Y$'s price with 26 tokens $X$ in our example visualized in Figure~\ref{fig:sandwich1}. The victim transaction $T_V$, which also buys asset $Y$, now only receives 52.205 tokens $Y$ -- almost 50\% less than expected -- and further inflates $Y$'s price. Finally, the attacker's back-running transaction sells the acquired asset $Y$ at a higher price. The attacker sells 68.286 tokens $Y$ for 58.017 tokens $X$, leaving the attacker with 32.017 additional tokens $X$. Thereby the sandwich attack increases the token's $Y$ price for the victim and provides the attacker with a net profit by taking advantage of the price curve's convexity. We note that the preceding example utilized extreme values to visualize the sandwich attack mechanism. Traders would generally set the slippage tolerance such that price movements of almost 50\% would lead to automatic trade failures and not allow for profitable sandwich attacks. Note that sandwich attacks are performed on individual trades as to push the individual trade to the maximum acceptable price movement. We further mention that sandwich attacks can also be similarly performed by liquidity providers. Liquidity providers front-run the victim's transaction by removing their liquidity and back-run it by adding the liquidity and profiting from the price change. In both cases, the profitability of a sandwich attack increases with the transaction size and slippage tolerance.

\begin{figure}[t]
    \centering    
    \usetikzlibrary{arrows}
\thispagestyle{empty}
\makeatletter
\pgfkeys{/kiviat/label style/.style={align=left,anchor=180+360/\tkz@kiv@radial*\rang}}

\usetikzlibrary{decorations.pathreplacing, arrows, fit}

\definecolor{color1}{HTML}{8EB1C7}
\definecolor{color2}{HTML}{FFBF69}
\definecolor{color3}{HTML}{B02E0C}
\definecolor{color4}{HTML}{B744B8}
\definecolor{color5}{HTML}{610345}
\definecolor{color6}{HTML}{044B7F}
\definecolor{color7}{HTML}{8E6C88}

\begin{tikzpicture}[scale = 1.5]

    \def \n {1.5}
	\def \radius {1cm}
	\def \margin {15} 

	\node[draw, thick,circle, minimum size = 1.3cm, fill =color1!50] at ({360/\n *3 +90}:\radius) (v5)  {\small X};
	\node[draw, thick,circle, minimum size = 1.3cm, fill =color2!50] at ({360/\n *1+90}:\radius) (v2) {\small Y};
	\node[draw, thick,circle, minimum size = 1.3cm, fill =color4!50] at ({360/\n *2+90}:\radius) (v3) {\small Z};

    \path[-latex, draw]
    
    (v2) edge[ thick, bend left = 0] node[midway,below] {\small$P_{Y\rightarrow Z}$}(v3)
    

    (v3) edge[thick, bend left = 0] node[midway,left] {\small$P_{Z\rightarrow X}$}(v5)

    (v5) edge[thick]node[midway,right] {\small$P_{X\rightarrow Y}$}(v2);
    
    
    \end{tikzpicture}\vspace{0pt}
    \caption{Cyclic trade execution between tokens $X$, $Y$ and $Z$. A trade is indicated by an arrow between two tokens and the received price is noted on the edges.} \label{fig:cyclic}\vspace{0pt}
\end{figure}
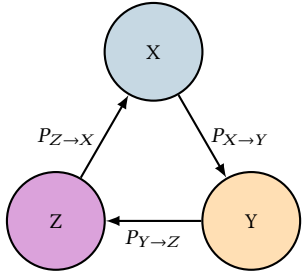

Further, CPMMs may present \emph{cyclic arbitrage opportunities} created by temporary price inaccuracies across liquidity pools~\cite{wang2022cyclic}. These arbitrage opportunities allow users to execute a cyclic trade with a net profit. For example, consider a CPMM with the following three liquidity pools between assets $X$, $Y$ and $Z$: $X\rightleftharpoons  Y$, $X\rightleftharpoons  Z$ and $Y\rightleftharpoons Z$ (cf. Figure~\ref{fig:cyclic}). When the price across pools is unsynced, if it is possible for a trader to trade tokens $X$ for tokens $Y$ in pool $X\rightleftharpoons  Y$ at the price $P_{X\rightarrow Y}$, then exchange tokens $Y$ for tokens $Z$ in pool  $Y\rightleftharpoons  Z$ at the price $P_{Y\rightarrow Z}$ and finally tokens $Z$ for tokens $X$ in pool $X\rightleftharpoons  Z$ at the price $P_{Z\rightarrow X}$ for a net profit. Note that the preceeding prices are the prices recevied by the trader and thus include the pool's transaction fee. The transaction is profitable, when 
$$P_{X\rightarrow Y}\cdot P_{Y\rightarrow Z}\cdot P_{Z\rightarrow X} \geq 1,$$
and allows the trader to extract more tokens $X$ from the last trade then initially enterted into the first trade.
When a user finds such an opportunity and submits a corresponding transaction to the mempool, anyone listening to the mempool can subsequently see this arbitrage opportunity. Through fatal front-running (cf. Figure~\ref{fig:fatalfrontrun}), an attacker can steal such an arbitrage opportunity. A similar transaction reordering attack has the attacker back-running (cf. Figure~\ref{fig:backrun}) the CPMM transaction that creates a market imbalance, thereby giving rise to an arbitrage opportunity. The attacker's back-running transaction then collects the arbitrage.

\subsection{Lending Protocols}
Lending protocols offer cryptocurrency loans in a trustless manner. Similar to DEXes, they are susceptible to transaction reordering attacks. In the following, we outline their general mechanism. Any user can participate as a lender by providing cryptocurrency assets to the protocol's smart contracts. For this service, lenders receive interest from the protocol's borrowers. Borrowers, on the other hand, must deposit collateral to take out a cryptocurrency loan and are charged periodically in the form of an interest rate. To allow for trustless loans and simultaneously protect the lender, the maximum loan value must be inferior to the borrower's collateral value. In this case, the loan is considered over-collateralized. If the value of the collateral drops below a pre-specified threshold, the protocol makes the loan available for liquidation~\cite{qin2021empirical}. Note that lending protocols utilize price oracles~\cite{liu2021first,eskandari2021sok} to determine the relative value of the collateral to the loan. Thus, the lending protocol only updates the relative value whenever the oracle updates. 

Two liquidation mechanisms dominate: (1) \emph{auction liquidation} and (2) \emph{fixed spread liquidation}. MakerDAO~\cite{2021makerdao}, the first platform to enable lending on the blockchain, utilizes auction liquidation. Once a loan becomes available for liquidation, interested liquidators can provide their bids to receive the loan's collateral. When the auction ends, the liquidator with the highest bid wins and receives the collateral. Aave~\cite{2021aave}, Compound~\cite{2021compound}, and dXdY~\cite{2021dxdy}, the largest lending protocols besides MarkerDAO, utilize fixed spread liquidations. Fixed spread liquidations make liquidated loans available immediately at a pre-determined discount. Liquidators assess whether liquidating the loan is profitable. The first liquidator engaging with a (profitable) liquidation opportunity then receives the collateral and claims the discount. When discussing BEV opportunities, we will focus on fixed spread liquidations. 

There are two common transaction reordering attacks on lending platforms utilizing fixed spread liquidations. In the first attack, the attacker observes an upcoming oracle update that will make a loan available for liquidation and back-runs (cf. Figure~\ref{fig:backrun}) the corresponding transaction. The attacker wants to place its transaction immediately after the oracle update transaction to be first to claim the discount. Additionally, an attacker might observe a liquidator attempting a profitable liquidation in the mempool and perform a fatal front-running transaction (cf. Figure~\ref{fig:fatalfrontrun}) to steal the associated profit.

\section{Fair Transaction Orderings and Measures for Mitigation Strategies}
    We start by noting that in terms of transaction reordering attack mitigation, a transaction order is considered \emph{fair} when it is not possible for any party to include or exclude transactions after seeing their contents. Further, it should not be possible for any party to insert their own transaction before any transaction whose contents it already been observed.

    In the following, we introduce seven measures with which we will assess approaches that mitigate transaction reordering manipulations and thereby enforce fair orderings and mitigate BEV:
    \paragraph{Decentralization} Decentralization measures the approach's impact on the blockchain's decentralization. We consider the Ethereum blockchain design, with miners building the blocks, as the benchmark and only assess whether and how significantly the approach decreases the level of decentralization.
    \paragraph{Security} Security measures how susceptible the approach is to attacks, i.e., how easily attackers bypass the BEV protection.
    \paragraph{Scope} Scope measures how wide-reaching the approach is, i.e., to what degree it prevents the varying BEV opportunities.
    \paragraph{Jostling} Jostling measures the competition between traders for block inclusion (at a preferential position). Jostling can arise from the approach limiting the transactions (per application) included in the block or (when possible) from traders over-bidding each other in a PGA. Both scenarios would lead traders to compete for a (preferential) spot in the block. Here, the Ethereum blockchain design serves as the benchmark for the middleground. 
    \paragraph{Goodput} Goodput measures whether the approach impacts the number of genuine transactions processed by the application or blockchain per time unit. We consider a transaction genuine when it is not part of an attack or an additional transaction required by the user to protect against an attack. As with decentralization, we consider the Ethereum blockchain design as the benchmark and assess to what degree the approach lessens the application's or blockchain's goodput. 
    \paragraph{Delay} Delay measures the time between a trade submission and its execution. Here, the Ethereum blockchain design also serves as the benchmark.
    \paragraph{Cost} Cost measures the additional cost the approach places directly on traders for executing their transaction. The additional costs could either be the additional gas cost stemming from transactions requiring more space on-chain or separate fees paid to those in charge of ordering the transactions.\\
    
    Rather than focusing solely on whether an approach successfully mitigates BEV, we aim to reflect its full impact on the blockchain with our measures. In the following, we will categorize BEV mitigation approaches and assess them in each measure with a score between 1 and 3. The lowest score -- 1 -- indicates that the approach performs poorly in a given measure. Satisfactory performance is indicated by the middle score -- 2. Finally, an approach that displays excellent performance a category is awarded the highest score -- 3.

\section{Mitigating Transaction Reordering Manipulations}
We categorize state-of-the-art mitigation approaches (operational projects, informal ideas, and academic works)  for transaction reordering manipulations in the following. To summarize each approach category, we assess it with our previously introduced measures allowing us to point out both its weaknesses and strengths.

\subsection{Optimized Trade Execution}\label{sec:optimizedtradeexecution}

We commence our analysis with the simplest and easiest to adopt BEV mitigation schemes. This category of schemes performs an application-specific transaction optimizations to mitigate specific attacks. 

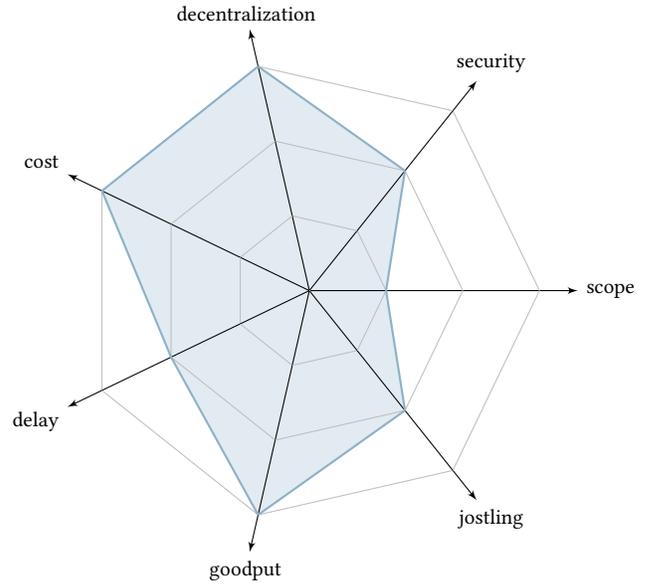
\begin{figure}[t]
    \centering    
    \usetikzlibrary{arrows}
\thispagestyle{empty}
\makeatletter
\pgfkeys{/kiviat/label style/.style={align=left,anchor=180+360/\tkz@kiv@radial*\rang}}

\usetikzlibrary{decorations.pathreplacing, arrows, fit}

\definecolor{color1}{HTML}{8EB1C7}
\definecolor{color2}{HTML}{FFBF69}
\definecolor{color3}{HTML}{B02E0C}
\definecolor{color4}{HTML}{B744B8}
\definecolor{color5}{HTML}{610345}
\definecolor{color6}{HTML}{044B7F}
\definecolor{color7}{HTML}{8E6C88}

\begin{tikzpicture}[]
\tikzstyle{every node}=[font=\small]

\tkzKiviatDiagram[scale=1.02,label space=0.5,
        radial  = 5,
        gap     = 1,  
        lattice = 3]{scope, security,decentralization,cost,delay,goodput, jostling}
\tkzKiviatLine[thick,color=color1,mark=none,
               fill=color1!50,opacity=.5](1,2,3,3,2,3,2)

\end{tikzpicture}\vspace{0pt}
    \caption{Optimized trade execution assessment. The approach's performance in each category is visualized in the spyder web. The spyder web's inner level represents the lowest score -- 1 -- and the web's outer layer corresponds to the highest score -- 3.} \label{fig:optimizedtradeexecution}\vspace{0pt}
\end{figure}

Zhou et al.~\cite{zhou2021a2mm} propose A2MM, an application that collects the created arbitrage opportunity directly. More specifically, the proposed scheme automatically checks whether a user's CPMM transaction would create an arbitrage opportunity, i.e., cause a market imbalance. For example, a relatively large trade might create a market imbalance causing a cyclic arbitrage opportunity. If so, the application proposed by Zhou et al. automatically collects the arbitrage within the same transaction and thereby does not leave any BEV behind. Note that the trader receives the collected arbitrage. Focusing on reducing sandwich attacks, Züst~\cite{zust2021analyzing} checks whether a user transaction is vulnerable to a sandwich attack. In case a profitable sandwich attack exists, generally the case for a relatively large transaction, the transaction is split into smaller unattackable transactions. Thereby, the scheme avoids sandwich attacks. In a similar line of work, Heimbach and Wattenhofer~\cite{heimbach2022eliminating} propose an algorithm that sets a CPMM transaction's slippage tolerance. The mechanism allows users to avoid both sandwich attacks and transaction failures from natural price movements. Thereby the scheme comes at a low cost for the user.

We assess the impact of optimized trade execution schemes on the blockchain in Figure~\ref{fig:optimizedtradeexecution}. The approach does not impact the system's decentralization and does not increase transaction costs. Note that \cite{zhou2021a2mm} can submit additional transactions to collect arbitrage. However, the collected arbitrage should exceed the additional fees. The same holds for \cite{zust2021analyzing}. Here, the additional transaction fees needed when splitting a transaction into multiple transactions must not top the user's cost associated with the potential sandwich attack. Decentralization is not impacted as the user simply changes the parameters relevant to their transactions to avoid attacks, and schemes do not impact how transactions are ordered. Further, the schemes leave the blockchain's goodput unchanged. We do not foresee a significant increase in unnatural transactions, and we expect fewer BEV collecting transactions. The additional transactions to collect the arbitrage in \cite{zhou2021a2mm} for instance, would otherwise be executed by arbitrageurs. Additional transactions stemming from splitting large trades in~\cite{zust2021analyzing} avoid two sandwich attack transactions taking up space on the blockchain, and we predict that very few trades will not be split more than three trades. Finally, the slippage setting algorithm presented in~\cite{heimbach2022eliminating} reduces both sandwich attacks and transaction failures. 

We foresee the potential for these schemes to increase jostling among similar transactions. The optimized user transactions are slightly less flexible than previously to protect users against BEV. Thus, the likelihood of two ordinary CPMM transactions that utilize the same protection mechanism competing for successful execution in the block is slightly elevated due to their increased complexity. On a similar note, we only expect a slight transaction delay increase. While transactions split by the mechanism suggested in \cite{zust2021analyzing} might execute over several blocks, creating a small delay, the reduced flexibility of the optimized transactions, in general, might lead to infrequent transaction failures causing a related delay. The approach's security against BEV mitigation has loopholes stemming from the blockchain's unknown state at execution time. For example, transactions deemed unattackable might become attackable due to a sudden drop in the gas fee. While such scenarios might occur in rare cases, most transactions will be successfully protected from the targeted attack. 

We conclude by noting that the most significant downfall of optimized trade execution schemes is their limited scope. The approach's scope is limited to particular attacks on specific applications. While optimized trade execution schemes cannot tackle general transaction reordering manipulations, they are well-suited to act as temporary solutions that users can apply themselves until a universal scheme preventing transaction reordering manipulations is implemented for their protection.

\subsection{Professional Market Makers}\label{sec:redesignDEX}



There are suggestions that go a step further than optimized trade execution by redesigning DEXes. Instead of utilizing an AMM these approaches introduce professional market makers. Their DEX redesign must automatically enforce fair orderings of DEX transactions on the blockchain. In the following, we discuss those that do not interfere with the blockchain's public nature.

Ciampi et al.~\cite{ciampi2021fairmm} introduce FairMM, which utilizes a monopolistic profit-seeking market maker to ensure a fair ordering. In the protocol, buyer $B_i$ first creates a smart contract $S_i$ and locks capital (token $X$) in the contract. Then she sends a trading request to buy token $Y$ to the single market maker (seller). The buyer sends the request off-chain, and the seller replies off-chain by proposing an exchange rate. Once the seller replies, the transaction order is locked and can no longer be hampered with. If the buyer agrees with the proposed exchange rate, she sends a certificate to the seller off-chain to buy $y$ tokens $Y$. This certificate can be used by the seller to withdraw the corresponding (according to the agreed-upon exchange rate) $x$ tokens $X$ from the smart contract $S_i$. Thereby, $S_i$ ensures that the sellers have send $y$ tokens $Y$ to the buyer $B_i$. Thus, FairMM does not function as an AMM but utilizes a single market maker. Hashflow~\cite{2022hashflow} is a DEX that takes a similar approach to redesign DEXes. Instead of introducing a monopolistic market maker, Hashflow invites professional market makers to manage liquidity pools and operates in the \emph{request-for-quote (RFQ)} model. In the RFQ model, users request a quote for their transaction off-chain from a professional market maker. The market maker replies by sending the user a signature-based quote off-chain. This quote allows the user to execute the trade at the offered price by broadcasting the transaction across the Ethereum network. As the price is agreed upon by the market maker and the buyer, the transaction will execute at this price. Further, Hashflow integrates a router that will execute your order on Uniswap instead if the price is better. 

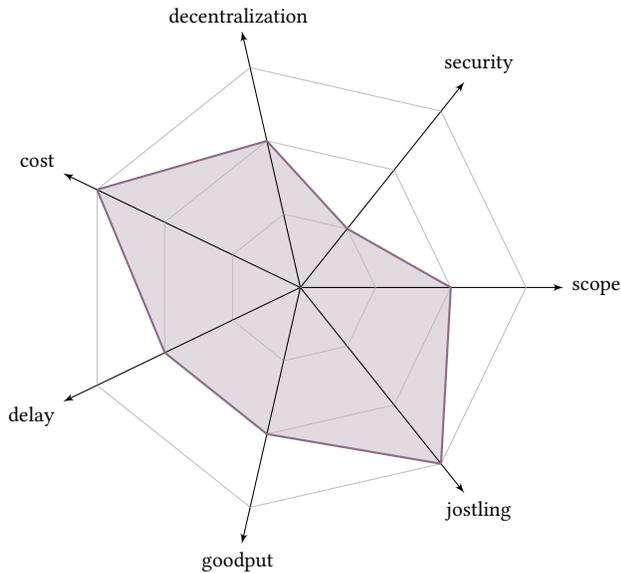
\begin{figure}[t]
    \centering    
    \usetikzlibrary{arrows}
\thispagestyle{empty}
\makeatletter
\pgfkeys{/kiviat/label style/.style={align=left,anchor=180+360/\tkz@kiv@radial*\rang}}

\usetikzlibrary{decorations.pathreplacing, arrows, fit}

\definecolor{color1}{HTML}{8EB1C7}
\definecolor{color2}{HTML}{FFBF69}
\definecolor{color3}{HTML}{B02E0C}
\definecolor{color4}{HTML}{B744B8}
\definecolor{color5}{HTML}{610345}
\definecolor{color6}{HTML}{044B7F}
\definecolor{color7}{HTML}{8E6C88}

\begin{tikzpicture}[]
\tikzstyle{every node}=[font=\small]

\tkzKiviatDiagram[scale=1,label space=0.5,
        radial  = 5,
        gap     = 1,  
        lattice = 3]{scope, security,decentralization,cost,delay,goodput, jostling}
\tkzKiviatLine[thick,color=color7,mark=none,
               fill=color7!50,opacity=.5](2,1,2,3,2,2,3)
               
\end{tikzpicture}\vspace{0pt}
    \caption{Professional market makers assessment. The approach's performance in each category is visualized in the spyder web. The spyder web's inner level represents the lowest score -- 1 -- and the web's outer layer corresponds to the highest score -- 3.} \label{fig:redesignDEX}
\end{figure}

In Figure~\ref{fig:redesignDEX} we provide an assessment of the professional market maker approach. The approach demonstrates average performance in most areas. As the approach focuses on DEXes, in particular, the scope is limited to transaction reordering manipulations on DEXes. We also anticipate the related and additional transactions to decrease goodput slightly. Small delay increases are very likely, as traders potentially first have to lock capital in a smart contract, then go through three rounds of communication and finally wait for the trade to execute on-chain when using FairMM. When using Hashflow, on the other hand, we only expect small delay increases stemming from off-chain communication in the RFQ model. 

We further do not expect a significant increase in cost. FairMM has buyers lock capital in the smart contract in a first transaction, and a second on-chain transaction is required for the funds to be exchanged. Note that the buyer can lock capital in the smart contract for several transactions. We, however, do expect that buyers will only lock capital in the smart contract for a small number of orders at once. In reality, many buyers will not be able to foresee all their future trades or might not have the available capital to lock for a large set of future trades. Hashflow, on the other hand, is more gas-efficient than Uniswap and, thus, might potentially decrease costs. In terms of jostling, we foresee excellent performance. Other traders do not observe the off-chain communication, and the ordering is set after the initial round of communication in both designs. Thus, we do not anticipate increased competition for a preferential position in a block. Decentralization, however, is impacted by having professional market makers. The professional market makers, at least to some extent, can control the ordering -- by choosing the order in which they reply to buyers' requests initially. Note that Hashflow allows for everyone to become a market maker, but one either requires significant capital or reputation (to receive capital from other users) in reality. Thus, market making is in the hands of professionals on Hashflow. 

Security is the approach's biggest downfall. On Hashflow, a transaction cannot be front-run once the trader has received a quote from the professional market maker when using Hashflow. However, with knowledge of the user's trade, the professional market maker could reply to their own front-running transaction before they reply to the user's transaction. Further, the professional market maker could front-run the trader's transaction on a centralized exchange, for instance. For FairMM, the approach's security largely relies on assuming rational behavior and is justified with associated economic models. We, however, expect that a byzantine market maker can attack the protocol. The market maker not only knows what, i.e., the specific cryptocurrencies, the trader wants to trade before the ordering is set but can also observe how much the trader has locked in the smart contract. We note here that then the blockchain's public nature allows the market maker to predict the user's trade size. Thus, there appears to be significant potential for attacks from the single market maker to attack. We further, which is disregarded in our assessment in Figure~\ref{fig:redesignDEX}, question the feasibility of the protocol, as the market maker cannot quote a price without knowing the trade size. FairMM expects that the market maker quotes a price without knowing the transaction's size. The price received in a trade execution in markets utilizing the limit order book mechanism or in CPMMs decreases with the trade size due to the market's limited depth. The effects of the limited market depth are, thus, disregarded by FairMM. Thus, we do not think of professional market makers as a flawless solution to transaction reordering attacks on the blockchain.

\subsection{Trusted Third Party Ordering}\label{sec:privateordering}

We observe the adoption of transaction reordering manipulation mitigation schemes utilizing trusted third parties occurring at the quickest pace. Similar to the previous approach, transactions are no longer broadcasted through Ethereum's peer-to-peer network. Instead, users send their transactions to the trusted third party tasked with ordering transactions, unlike in the previous approach where the party is responsible for market making. 

Services such as flashbots~\cite{2021flash}, Eden~\cite{2021eden}, and OpenMEV~\cite{2022openmev} are already widely adopted on the Ethereum blockchain and rely on a trusted third party to order transactions. Users can send their transactions to these services without previously broadcasting them across the peer-to-peer network. Ordered transaction bundles from flashbots, Eden, and OpenMEV are send directly to miners for block inclusion. Thus, users rely on several trusted third parties, the service, and the miner to order their transactions. While users have to actively seek out these services for front-running protection on most DEXes, some DEXes have integrated the services directly. While SushiSwap~\cite{2021sushiswap} allows users to send their transactions through Eden in their API, mistX~\cite{2022mistx} is a DEX that integrates flashbots for all transactions. 

There is also talk of the adaption of a similar private ordering approach in ETH 2.0~\cite{buterin2021proposer}, whereby users no longer broadcast transactions to the mempool but instead directly send their transactions to a block proposer. The block proposer then submits an ordered transaction bundle to the block's miner/validator. Transactions ordered by a trusted third party (block proposer) do not enter the public mempool before their execution, and, therefore, they cannot be front-run if the trusted third party is honest. Gnosis Protocol~\cite{2022gnosis}, otherwise known as CowSwap~\cite{2022cowswap}, is a price discovery mechanism for DEXes that functions similarly. In the Gnosis Protocol, users send their transactions directly to \emph{solvers}. Solvers are given a similar role as block proposers. In particular, it is their job to provide settlement solutions for a batch of transactions. Note that batches are of limited size, and a uniform clearing price must be enforced, i.e., all orders receive assets priced equally against each other. Multiple solvers compete with each other to provide the best batch and are rewarded when they find the best one. As all transactions within a batch execute at the same price, there can be no sandwich attacks within a batch. 

Bentov et al.~\cite{bentov2019tesseract} introduce Tesseract: a real-time exchange that relies on trusted hardware to resit front-running. Tesseract operates through a \emph{trusted execution environment (TEE)}~\cite{zhang2016sok,pass2017formal} in which code is neither observed nor can it be tampered with. The trusted hardware orders transactions according to their arrival times. Tesseract requires trust, as it relies on a trusted third party: the hardware manufacturer. In a similar line of work, Stathakopoulou et al.~\cite{stathakopoulou2021adding} present Fairy to bring fairness to ordering. Fairy implements a fair ordering by relying on a TEE. 

Protocols utilitzing TEE are already deployed across DeFi. Automata~\cite{2022ata} utilizes a TEE to implement a conveyor service that determines the order of incoming transactions and thereby creates a front-running free zone. XATA~\cite{2022XATA} is a DEX that relies on Automata's conveyor service and only accepts transactions in the conveyor's pre-defined order. Similarly, SecretSwap~\cite{2022secretswap} relies on Secret Network~\cite{2022secretnetwork}, which processes transactions privately in a TEE, to implement a front-running resistant DEX. Trades on SecretSwap execute secretly. Thus, SecretSwap also preserves a transaction's privacy after its execution, at least to an extent. As the liquidity pool's state is public, one can infer an individual trades size, especially when the volume is small. 

\begin{figure}[t]
    \centering    
    \usetikzlibrary{arrows}
\thispagestyle{empty}
\makeatletter
\pgfkeys{/kiviat/label style/.style={align=left,anchor=180+360/\tkz@kiv@radial*\rang}}

\usetikzlibrary{decorations.pathreplacing, arrows, fit}

\definecolor{color1}{HTML}{8EB1C7}
\definecolor{color2}{HTML}{FFBF69}
\definecolor{color3}{HTML}{B02E0C}
\definecolor{color4}{HTML}{B744B8}
\definecolor{color5}{HTML}{610345}
\definecolor{color6}{HTML}{044B7F}
\definecolor{color7}{HTML}{8E6C88}

\begin{tikzpicture}[]
\tikzstyle{every node}=[font=\small]

\tkzKiviatDiagram[scale=1.02,label space=0.5,
        radial  = 5,
        gap     = 1,  
        lattice = 3]{scope, security,decentralization,cost,delay,goodput, jostling}

\tkzKiviatLine[thick,color=color2,mark=none,
               fill=color2!50,opacity=.5](3,1,1,2,3,3,3)
               
\end{tikzpicture}\vspace{0pt}
    \caption{Trusted third party ordering assessment. The approach's performance in each category is visualized in the spyder web. The spyder web's inner level represents the lowest score -- 1 -- and the web's outer layer corresponds to the highest score -- 3.} \label{fig:privateordering}
\end{figure}
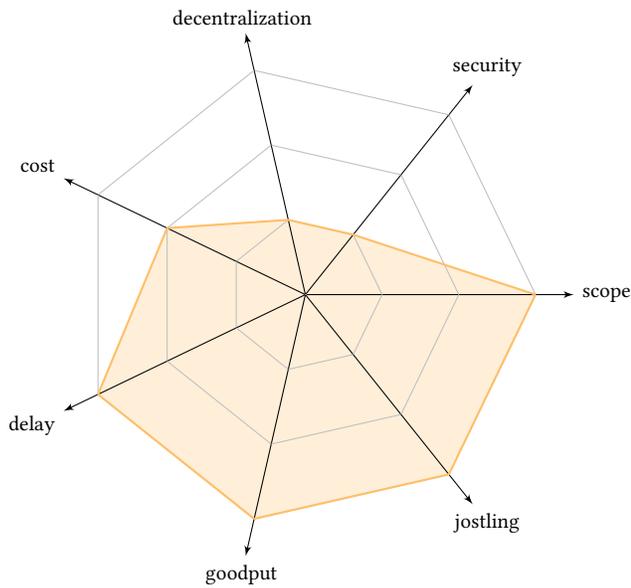

In Figure~\ref{fig:privateordering}, we assess trusted third party ordering as a BEV mitigation scheme. Trusted third party ordering exhibits a very asymmetric performance across our seven measures. The approach excels in four of our seven measures: delay, goodput, jostling, and scope. For one, the transaction fee should not increase, as the transaction does not require more space on-chain. There is no reason to believe that ordering by a trusted third party increases delay. Transactions simply wait with the trusted third party until they can be included in a block. Otherwise, they would wait in the mempool for block inclusion. The approach further is not expected to impact goodput negatively, as the trusted third party sometimes even ensures that transactions doomed for failure are never included in the block~\cite{2021flash,2021eden,2022openmev}, and further fewer transactions are visible for attacks. Jostling is also expected to be very low due to the approach's private nature. The strain PGAs, caused by arbitrageurs competing for a BEV opportunity, place on the network's gas price is reduced. We also note that the approach's scope is broad as it has the potential to mitigate all kinds of transaction reordering manipulations related to BEV. In terms of costs, we anticipate a mediocre performance. At the moment, trusted third parties generally charge little to nothing for their ordering service, but we expect these costs to increase. Further, setup costs for trusted hardware might be significant.

The remaining measures, decentralization, and security, are where private ordering falls short. Ordering responsibility is placed in the hands of a trusted third party, not only making ordering centralized but also relying solely on the honest behavior of the trusted third party for security. A byzantine trusted third party could, therefore, easily manipulate transaction orderings. To summarize, the approach relies on centralized ordering. Thereby gaining significant performance benefits similar to those observed in centralized exchanges, but is in stark contradiction to the blockchain's trustless nature.

\subsection{eUTXO Model}\label{sec:eutxo}

Imagine a transaction that only executes on the state that it sees at the time of submission. Such a transaction can not be front-run as the attacker cannot manipulate the state to their advantage without inducing the transaction's automatic failure. Utilizing the eUTXO model to mitigate transaction reordering manipulations builds on this idea. 

Lanningham~\cite{lanningham2021sundaeswap} presented a translation of the Uniswap model to Cardano's~\cite{david2018ouroboros} \emph{extended UTXO (eUTXO)} model~\cite{chakravarty2020extended} in the SundaeSwap~\cite{2021sundaeswap} whitepaper. As the name suggests, the eUTXO model is an extension of Bitcoin's \emph{unspent transaction output (UTXO)} model, whereby a user's coins are not accumulated in their account but instead stored in individual UTXOs linked to their account. Each transaction takes UTXOs as inputs, destroys them in the process, and outputs a new set of UTXOs. The translation presented from the Uniswap model to Cardano in \cite{lanningham2021sundaeswap} stores a liquidity pool's liquidity in an eUTXO, thus, simulating a smart contract with an eUTXO. When submitting a DEX transaction, users need to interact with the specific liquidity pool's eUTXO. Each time a user interacts with the pool's eUTXO, a new eUTXO is created in its place while the old one is destroyed. Thus, only a single interaction with the pool per block is possible. The limited transaction throughput stems from the fact that users must reference the pool's eUTXO in their transaction, which is no longer valid as soon as a single transaction is executed. This straightforward adaption of Uniswap's AMM design would therefore prevent front-running -- mitigating BEV -- as no user transaction can be forced to execute on a new state. A similar implementation of a DEX on the Ethereum chain could also set the slippage tolerance of all transactions to zero or by only letting one transaction interact with the smart contract per block. 

\begin{figure}[t]
    \centering    
    \usetikzlibrary{arrows}
\thispagestyle{empty}
\makeatletter
\pgfkeys{/kiviat/label style/.style={align=left,anchor=180+360/\tkz@kiv@radial*\rang}}

\usetikzlibrary{decorations.pathreplacing, arrows, fit}

\definecolor{color1}{HTML}{8EB1C7}
\definecolor{color2}{HTML}{FFBF69}
\definecolor{color3}{HTML}{B02E0C}
\definecolor{color4}{HTML}{B744B8}
\definecolor{color5}{HTML}{610345}
\definecolor{color6}{HTML}{044B7F}
\definecolor{color7}{HTML}{8E6C88}

\begin{tikzpicture}[]
\tikzstyle{every node}=[font=\small]

\tkzKiviatDiagram[scale=1.02,label space=0.5,
        radial  = 5,
        gap     = 1,  
        lattice = 3]{scope, security,decentralization,cost,delay,goodput, jostling}
\tkzKiviatLine[thick,color=color3,mark=none,
               fill=color3!50,opacity=.5](1,3,3,3,1,1,1)

\end{tikzpicture}\vspace{0pt}
    \caption{eUTXO model assesment. The approach's performance in each category is visualized in the spyder web. The spyder web's inner level represents the lowest score -- 1 -- and the web's outer layer corresponds to the highest score -- 3.} \label{fig:eutxo}\vspace{0pt}
\end{figure}
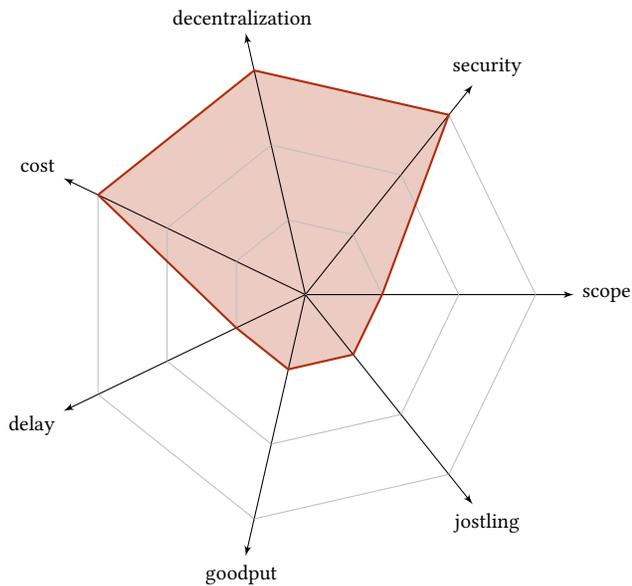

In our assessment (cf. Figure~\ref{fig:eutxo}), we find that the approach does not reduce the blockchain's decentralization and is very secure, as there is no possibility to be front-run and forced to execute with an unwanted state. Similarly, the transaction fee should not increase as each transaction does not require more space on-chain. Concerning the scope of the approach, it prevents front-running BEV but does not prevent fatal front-running and potentially back-running. Thus, the scope is not significantly more general than that of the individual optimized trade execution approaches presented in Section~\ref{sec:optimizedtradeexecution}. The main problem of utilizing the eUTXO model to prevent BEV is easily identified. The approach's goodput is extremely limited, as it only allows for a single transaction per block in a liquidity pool. This bottleneck also leads to significant jostling, with all transactions in a liquidity pool competing for the same spot for block inclusion. Naturally, the delay is also impacted, especially in liquidity pools between more popular cryptocurrencies. 

Thus, regardless of the successful BEV mitigation, such a DEX would not be equipped to handle the goodput demands placed on it, and this explains why SundaeSwap alerted its design before it ever launched~\cite{2022meet}.

\subsection{Algorithmic Committee Ordering}\label{sec:committeeordering}

Several protocols relying on a committee to algorithmically order transactions fairly have surfaced recently. The general idea of the approach is for a committee to observe incoming transactions and to agree on a fair ordering through consensus. The main challenge presented to these protocols is that, as the Condorcet paradox~\cite{gehrlein1983condorcet} shows, even when all committee members are honest, they can not agree on a fair ordering~\cite{kelkar2020order}. Consider a committee composed of three members and a set of three transactions: $T_1$, $T_2$, and $T_3$. The transactions arrive at each committee member in the following orders: 
\begin{itemize}
    \item committee member one: $T_1$, $T_2$, $T_3$,
    \item committee member two: $T_2$, $T_3$, $T_1$,
    \item committee member three: $T_3$, $T_1$, $T_2$.
\end{itemize}
Thus, a majority of the committee members observe transaction $T_1$ before transaction $T_2$, transaction $T_2$ before transaction $T_3$, and finally transaction $T_3$ before transaction $T_1$. Therefore, it is impossible for a committee to agree on a fair ordering, resulting in the schemes generally introducing a weakened fairness definition. We note that while this field of research is not limited to fair orderings that mitigate BEV~\cite{lev2019fairledger,sokolik2020age}, we will solely focus on those protocols applicable to BEV mitigation. 

Baird~\cite{baird2016swirlds} introduces Hashgraph consensus that relies on a gossip protocol. Committee members gossip about gossip of their received transaction orders and utilize virtual voting to agree on a fair order such that no small group of attackers can unfairly influence the order of transactions. Virtual voting allows Hashgraph to reduce the system's complexity. However, as fairness is only vaguely defined in Hashgraph and therefore some questions regarding the fairness guarantees of the approach remain. Kurswae~\cite{kursawe2020wendy} presents a group of protocols that utilize Byzantine consensus to ensure \emph{relative order fairness}, whereby an ordering is fair if transaction $T$ that was seen by all honest nodes before transaction $T'$ then transaction $T$ executes before transaction $T'$. A similar fairness notion is achieved through the \emph{Byzantine ordered consensus} by Zhang et al.~\cite{zhang2020byzantine}. The fairness notion is strengthened in a recent line of work by Kelkar et al.~\cite{kelkar2020order,kelkar2021order,kelkar2021themis} that achieve \emph{$\gamma$-block-order-fairness}. If a $\gamma$ fraction of nodes receive transaction $T$ before transaction $T'$ then transaction $T$ executes before transaction $T'$. While \cite{kelkar2020order} and \cite{kelkar2021themis} utilize Byzantine consensus, the committee is permissionless in \cite{kelkar2021order}. Cachin et al.~\cite{cachin2021quick} even further strengthen the fairness notion by achieving \emph{$\bar{\gamma}$-block-order-fairness} via Byzantine consensus. In the previous, $\bar{\gamma}$ denote the fraction of correct nodes that receive transaction $T$ before transaction $T'$. 

\begin{figure}[t]
    \centering    
    \usetikzlibrary{arrows}
\thispagestyle{empty}
\makeatletter
\pgfkeys{/kiviat/label style/.style={align=left,anchor=180+360/\tkz@kiv@radial*\rang}}

\usetikzlibrary{decorations.pathreplacing, arrows, fit}

\definecolor{color1}{HTML}{8EB1C7}
\definecolor{color2}{HTML}{FFBF69}
\definecolor{color3}{HTML}{B02E0C}
\definecolor{color4}{HTML}{B744B8}
\definecolor{color5}{HTML}{610345}
\definecolor{color6}{HTML}{044B7F}
\definecolor{color7}{HTML}{8E6C88}

\begin{tikzpicture}[]
\tikzstyle{every node}=[font=\small]

\tkzKiviatDiagram[scale=1.02,label space=0.5,
        radial  = 5,
        gap     = 1,  
        lattice = 3]{scope, security,decentralization,cost,delay,goodput, jostling}
\tkzKiviatLine[thick,color=color4,mark=none,
               fill=color4!50,opacity=.5](2,1,2,2,3,3,2)

\end{tikzpicture}\vspace{0pt}
    \caption{Algorithmic committee ordering assessment. The approach's performance in each category is visualized in the spyder web. The spyder web's inner level represents the lowest score -- 1 -- and the web's outer layer corresponds to the highest score -- 3.} \label{fig:committeeordering}\vspace{0pt}
\end{figure}

Algorithmic committee ordering protocols are reaching ever-increasing levels of fairness when assessing them by the order in which nodes receive transactions. However, this notion of fairness disregards the potential for attackers to have better network connections. The attacker can listen to incoming transactions in the network and (fatally) front-run these transactions by having better connections, i.e., the attacker's transaction reaches the majority of the honest nodes before the victim's transaction even though the attacker only acted upon seeing the victim transaction. Such races take place in traditional finance~\cite{daian2020flash}. Back-running, while generally less severe on its own, remains easily possible. Thereby the protocol, even if all committee nodes were honest, would not mitigate all transaction reordering manipulations -- letting the approach only achieve a medium score in terms of its scope (cf. Figure~\ref{fig:committeeordering}). The protocols generally rely on a two-thirds majority of honest nodes. We consider this assumption to be too strong in the crypto space, where players are generally rational, becoming ever more apparent given the levels of BEV extraction~\cite{qin2021quantifying,torres2021frontrunner,zhou2021just}. Therefore, the required trust and the remaining (fatal) front-running possibilities outlined previously lead us to conclude that the system is not secure against attacks. Algorithmic committee ordering reduces the system decentralization. In the majority of protocols, a permissioned committee is in charge of the ordering, while in the permissionless case~\cite{kelkar2021order} the nodes' incentives are unclear. Thus, the ordering is generally placed in the hands of a few -- increasing the blockchain's centralization. 

We, further, expect transactions to become slightly more expensive as we expect blockchain implementations to provide financial incentives to the committee. Note that implementing such an ordering on top of the blockchain might require additional space on-chain -- increasing transaction fees. While we foresee some jostling, given the potential for (fatal) front-running, we don't expect higher levels than in the current setting. Finally, we expect that implementations of the algorithmic committee ordering approach should match at least the current goodput levels. We do not obverse a reason for the approach to infer any significant delay in the system. The time taken by committee members to order transactions is unlikely to exceed the time needed to build blocks. 

Algorithmic committee ordering presents a middle ground between a trusted third party and a decentralized ordering. However, the possibility of better network connections owned by the attacker and the assumptions regarding the committee's honesty limit the potential of the algorithmic committee ordering.

\subsection{On-Chain Commit \& Reveal}\label{sec:onchainreveal}

On-chain commit \& reveal as opposed to its off-chain counterpart, which we will present in the proceeding Section (cf. Section~\ref{sec:offchainreveal}), orders transactions on-chain -- as suggested by the name. The approaches generally consist of two phases. In the first phase, users commit to their transactions, and then in the second phase, the users themselves or the chain automatically reveal the transaction in a later block. 

Tatabitovska et al.~\cite{tatabitovska2021mitigation} propose a Uniswap specific commit \& reveal scheme utilizing hash commitments -- an adjustment of LibSubmarine’s commitment which Breidenbach et al.~\cite{breidenbach2018enter} introduce. LibSubmarine commitments hide the pool's address by hashing it. The commit transaction is recorded on-chain and ordered for later execution. A second reveal transaction is then send after a minimum number of blocks. The actual Uniswap trade executes upon block inclusion of the reveal transaction. The authors either have the users themselves reveal their commitment or let Uniswap implement a queue that handles the reveal. We only consider the implementation where the users themselves are in charge of the reveal phase, as the approach would otherwise fall under trusted third party ordering (cf. Section~\ref{sec:privateordering}). 

Instead of using hash commitment, Doweck and Eyal~\cite{doweck2020multi} make use of \emph{time-locke puzzle commitments}~\cite{rivest1996time}. They introduce \emph{Multi-Party Timed Commitments (MPTC)} to formalize the problem of implementing commitments with a probabilistic delay and propose the \emph{Time Capsule} protocol to solve the problem. In their scheme, $n$ users jointly build a random public key. They each send a random nonce to the coordinator, who aggregates the nonce in a Merkle tree. The tree's root then serves as the random seed for the public key, and users send their transaction, encrypted with the public key by using El-gamal encryption, to the coordinator. The coordinator then attempts to decrypt the message, which succeeds after a predictable time and reveals the messages. 

\begin{figure}[t]
    \centering    
    \usetikzlibrary{arrows}
\thispagestyle{empty}
\makeatletter
\pgfkeys{/kiviat/label style/.style={align=left,anchor=180+360/\tkz@kiv@radial*\rang}}

\usetikzlibrary{decorations.pathreplacing, arrows, fit}

\definecolor{color1}{HTML}{8EB1C7}
\definecolor{color2}{HTML}{FFBF69}
\definecolor{color3}{HTML}{B02E0C}
\definecolor{color4}{HTML}{B744B8}
\definecolor{color5}{HTML}{610345}
\definecolor{color6}{HTML}{044B7F}
\definecolor{color7}{HTML}{8E6C88}

\begin{tikzpicture}[]
\tikzstyle{every node}=[font=\small]

\tkzKiviatDiagram[scale=1.02,label space=0.5,
        radial  = 5,
        gap     = 1,  
        lattice = 3]{scope, security,decentralization,cost,delay,goodput, jostling}
\tkzKiviatLine[thick,color=color5,mark=none,
               fill=color5!50,opacity=.5](3,2,3,1,1,2,3)
\end{tikzpicture}\vspace{-9pt}\vspace{-2pt}
    
    \caption{On-chain commit \& reveal assessment. The approach's performance in each category is visualized in the spyder web. The spyder web's inner level represents the lowest score -- 1 -- and the web's outer layer corresponds to the highest score -- 3.} \label{fig:onchainreveal}\vspace{0pt}
\end{figure}
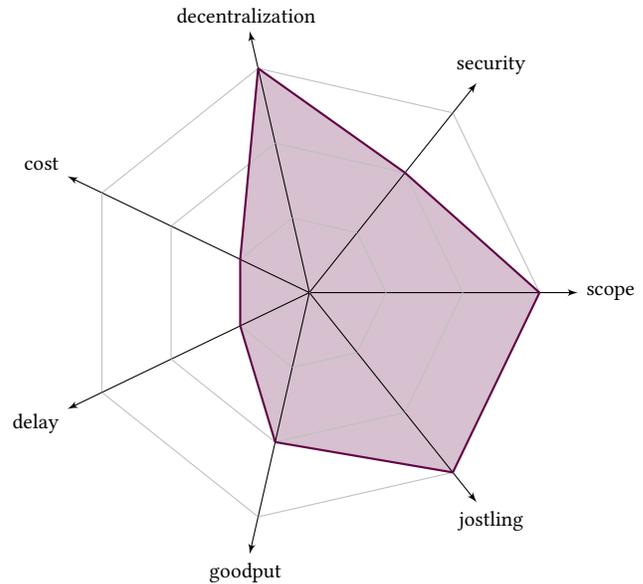

We find that on-chain commit \& reveal does not impact decentralization negatively (cf. Figure~\ref{fig:onchainreveal}). While the scheme presented by Doweck and Eyal utilizes a coordinator, even an adversarial coordinator cannot gain a significant advantage in solving the puzzle. Further, the approach tackles all transaction reordering manipulations, except for the adversarial committee member placing their own transaction first. As transaction contents are hidden, we expect jostling to be minimal. We further note that transactions which are unaware of each other are less likely to compete for the same spot in a block. On a similar note, we can only observe small security loopholes, such as an attacker choosing not to reveal its transaction in the scheme presented by Tatabitovska et al.~\cite{tatabitovska2021mitigation}. 

The commit and reveal transactions are included on-chain. Thus, significant additional space is required on-chain -- increasing the transaction fee. We, therefore, can only give the scheme a low score in our cost measure. The other major downfall is the delay between the transaction submission and execution. The approach adds a delay of a few block times between the commit transaction and the reveal transaction to the current delay. Especially when cryptocurrency prices are volatile, this additional delay can be detrimental to a protocol. For example, the price of Ether in US\$ changed by more than 2\% on 24 January 2022 at 8:47 (UTC$\pm$0)~\cite{2022ETHUSDT}. Thus, a DEX relying on on-chain commit \& reveal cannot reflect market prices accuracy, and market inefficiencies~\cite{malkiel1989stock}, which create cyclic arbitrage opportunities~\cite{Berg2022empirical}, are to be expected. We also note that the state of DEX, relying on on-chain commit \& reveal for ordering, is unknown to the user at the time of submission. 

To conclude, we expect to see an increase in transaction failures due to price movements. Increased transaction failures would reduce the DEXes goodput. While on-chain commit \& reveal might find a use case outside of DEXes and lending protocols, we do not see it meeting the delay and cost expectation placed on DEXes.

\subsection{Off-Chain Commit \& Reveal}\label{sec:offchainreveal}

Similar to on-chain commit \& reveal schemes, off-chain commit \& reveal schemes consist of a first round of traders committing to their transactions and a second round where the transactions are revealed. However, unlike their on-chain counterpart, the commitment does not take place on-chain but is instead handled off-chain by a committee. 

Reiter and Birman~\cite{reiter1994securely}, Miller et al.~\cite{miller2016honey} and Asyag et al.~\cite{asayag2018fair} introduce encrypted Byzantine consensus protocols relying on threshold signatures~\cite{shamir1979share,blakley1979safeguarding,desmedt1989threshold}. In an $(l,n)$ threshold signature scheme, a single public key is used to encrypt messages, and $l$ out of $n$ committee member signatures combine to decrypt the message. While the protocols introduced by Reiter and Birman, as well as Miller et al., are not aimed at preventing transaction reordering manipulations, the schemes can be utilized to achieve fair transaction orderings. The protocols have users encrypt their transactions with the public key, and then the committee orders encrypted transactions. A two-thirds majority of the committee is assumed to be honest. Upon agreeing on the final ordering, the committee members decrypt the messages together using their threshold signatures. Asyag et al. implement a similar ordering protocol and also make use of an additional randomness beacon to select the leader for the Byzantine consensus. 

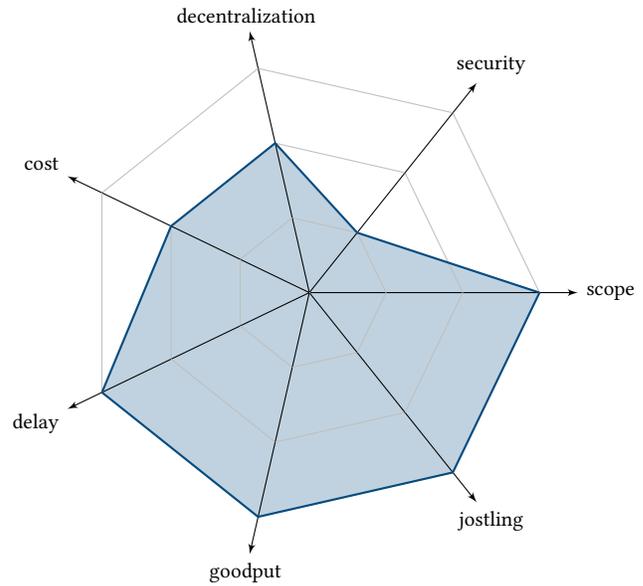
\begin{figure}[t]
    \centering    
    \usetikzlibrary{arrows}
\thispagestyle{empty}
\makeatletter
\pgfkeys{/kiviat/label style/.style={align=left,anchor=180+360/\tkz@kiv@radial*\rang}}

\usetikzlibrary{decorations.pathreplacing, arrows, fit}

\definecolor{color1}{HTML}{8EB1C7}
\definecolor{color2}{HTML}{FFBF69}
\definecolor{color3}{HTML}{B02E0C}
\definecolor{color4}{HTML}{B744B8}
\definecolor{color5}{HTML}{610345}
\definecolor{color6}{HTML}{044B7F}
\definecolor{color7}{HTML}{8E6C88}

\begin{tikzpicture}[]
\tikzstyle{every node}=[font=\small]

\tkzKiviatDiagram[scale=1.02,label space=0.5,
        radial  = 5,
        gap     = 1,  
        lattice = 3]{scope, security,decentralization,cost,delay,goodput, jostling}
\tkzKiviatLine[thick,color=color6,mark=none,
               fill=color6!50,opacity=.5](3,1,2,2,3,3,3)
               
\end{tikzpicture}\vspace{0pt}
    \caption{Off-chain commit \& reveal assessment. The approach's performance in each category is visualized in the spyder web. The spyder web's inner level represents the lowest score -- 1 -- and the web's outer layer corresponds to the highest score -- 3.} \label{fig:offchainreveal}\vspace{0pt}
\end{figure}

Similar to its on-chain counterpart (cf. Section~\ref{sec:onchainreveal}), off-chain commit \& reveal excels in terms of scope and jostling (cf. Figure~\ref{fig:offchainreveal}). All transaction reordering manipulation attacks, except for the attacker placing its own transaction first, are tackled. Additionally, as the transaction contents are not public before the ordering is fixed, we expect jostling to be reduced. In terms of cost, we predict transactions to become more expensive. As with algorithmic committee ordering (cf. Section~\ref{sec:committeeordering}), we do not foresee the approach impacting delay or goodput negatively. The committee is not expected to take longer to order a set of transactions than the time the Ethereum network requires to create a new block. Goodput is foreseen to stay at current levels: users do not have to submit additional transactions, and we expect the number of adversary transactions to decrease. We again expect blockchain implementations to introduce financial incentives for the committee members. Off-chain commit \& reveal blockchain implementations might require extra space on-chain, e.g., for the committee member signatures. Thus, we predict an increase in transaction fees. With ordering in the hands of a permissioned committee, the approach reduces the decentralization of ordering. This lack of decentralization and the committee's ability to perform arbitrary transaction reordering manipulations when colluding with each other is the reason for the approach's poor performance in terms of security. 

With the exception of decentralization and security, off-chain commit \& reveal appears to generally combine the benefits of algorithmic committee ordering (cf. Section~\ref{sec:committeeordering}) and on-chain commit \& reveal (cf. Section~\ref{sec:onchainreveal}). The approach's weaknesses -- decentralization and security -- are those most at odds with the blockchain's most fundamental principles. 

\section{Discussion}

\begin{figure}[t]
    \centering    
    \begin{tikzpicture}[scale = 0.83]

\node[anchor = east, align = right] at (0.45,-0.1)  [rotate=40,font=\small]
    {scope};
    \node[anchor = east, align = right] at (1.45,-0.1)  [rotate=40,font=\small]
    {security};
    \node[anchor = east, align = right] at (2.45,-0.1)  [rotate=40,font=\small]
    {decentralization};
    \node[anchor = east, align = right] at (3.45,-0.1)  [rotate=40,font=\small]
    {cost};
    \node[anchor = east, align = right] at (4.45,-0.1)  [rotate=40,font=\small]
    {delay};
    \node[anchor = east, align = right] at (5.45,-0.1)  [rotate=40,font=\small]
    {goodput};
    \node[anchor = east, align = right] at (6.45,-0.1)  [rotate=40,font=\small]
    {jostling};

 \node[text width=2.5cm,anchor = east] at (-0.02,0.45)  [font=\small ]
    {off-chain commit \& reveal};
\filldraw[fill=lightgreen, rounded corners = 2pt] (0,0) rectangle (0.9,0.9);
\filldraw[fill=lightred, rounded corners = 2pt] (1,0) rectangle (1.9,0.9);
\filldraw[fill=lightyellow, rounded corners = 2pt] (2,0) rectangle (2.9,0.9);
\filldraw[fill=lightyellow, rounded corners = 2pt] (3,0) rectangle (3.9,0.9);
\filldraw[fill=lightgreen, rounded corners = 2pt] (4,0) rectangle (4.9,0.9);
\filldraw[fill=lightgreen, rounded corners = 2pt] (5,0) rectangle (5.9,0.9);
\filldraw[fill=lightgreen, rounded corners = 2pt] (6,0) rectangle (6.9,0.9);

 \node[text width=2.5cm,anchor = east] at (-0.02,1.45)  [font=\small]
    { on-chain commit \& reveal};
\filldraw[fill=lightgreen, rounded corners = 2pt] (0,1) rectangle (0.9,1.9);
\filldraw[fill=lightyellow, rounded corners = 2pt] (1,1) rectangle (1.9,1.9);
\filldraw[fill=lightgreen, rounded corners = 2pt] (2,1) rectangle (2.9,1.9);
\filldraw[fill=lightred, rounded corners = 2pt] (3,1) rectangle (3.9,1.9);
\filldraw[fill=lightred, rounded corners = 2pt] (4,1) rectangle (4.9,1.9);
\filldraw[fill=lightyellow, rounded corners = 2pt] (5,1) rectangle (5.9,1.9);
\filldraw[fill=lightgreen, rounded corners = 2pt] (6,1) rectangle (6.9,1.9);

 \node[text width=2.5cm,anchor = east] at (-0.02,2.45)  [font=\small]
    {algorithmic committee ordering};
\filldraw[fill=lightyellow, rounded corners = 2pt] (0,2) rectangle (0.9,2.9);
\filldraw[fill=lightred, rounded corners = 2pt] (1,2) rectangle (1.9,2.9);
\filldraw[fill=lightyellow, rounded corners = 2pt] (2,2) rectangle (2.9,2.9);
\filldraw[fill=lightyellow, rounded corners = 2pt] (3,2) rectangle (3.9,2.9);
\filldraw[fill=lightgreen, rounded corners = 2pt] (4,2) rectangle (4.9,2.9);
\filldraw[fill=lightgreen, rounded corners = 2pt] (5,2) rectangle (5.9,2.9);
\filldraw[fill=lightyellow, rounded corners = 2pt] (6,2) rectangle (6.9,2.9);

 \node[text width=2.5cm,anchor = east] at (-0.02,3.45)  [font=\small]
    {  eUTXO model};
\filldraw[fill=lightred, rounded corners = 2pt] (0,3) rectangle (0.9,3.9);
\filldraw[fill=lightgreen, rounded corners = 2pt] (1,3) rectangle (1.9,3.9);
\filldraw[fill=lightgreen, rounded corners = 2pt] (2,3) rectangle (2.9,3.9);
\filldraw[fill=lightgreen, rounded corners = 2pt] (3,3) rectangle (3.9,3.9);
\filldraw[fill=lightred, rounded corners = 2pt] (4,3) rectangle (4.9,3.9);
\filldraw[fill=lightred, rounded corners = 2pt] (5,3) rectangle (5.9,3.9);
\filldraw[fill=lightred, rounded corners = 2pt] (6,3) rectangle (6.9,3.9);

 \node[text width=2.5cm,anchor = east] at (-0.02,4.45)  [font=\small]
    {  trusted third party ordering};
\filldraw[fill=lightgreen, rounded corners = 2pt] (0,4) rectangle (0.9,4.9);
\filldraw[fill=lightred, rounded corners = 2pt] (1,4) rectangle (1.9,4.9);
\filldraw[fill=lightred, rounded corners = 2pt] (2,4) rectangle (2.9,4.9);
\filldraw[fill=lightyellow, rounded corners = 2pt] (3,4) rectangle (3.9,4.9);
\filldraw[fill=lightgreen, rounded corners = 2pt] (4,4) rectangle (4.9,4.9);
\filldraw[fill=lightgreen, rounded corners = 2pt] (5,4) rectangle (5.9,4.9);
\filldraw[fill=lightgreen, rounded corners = 2pt] (6,4) rectangle (6.9,4.9);

 \node[text width=2.5cm,anchor = east] at (-0.02,5.45)  [font=\small]
    { professional market makers};
\filldraw[fill=lightyellow, rounded corners = 2pt] (0,5) rectangle (0.9,5.9);
\filldraw[fill=lightred, rounded corners = 2pt] (1,5) rectangle (1.9,5.9);
\filldraw[fill=lightyellow, rounded corners = 2pt] (2,5) rectangle (2.9,5.9);
\filldraw[fill=lightgreen, rounded corners = 2pt] (3,5) rectangle (3.9,5.9);
\filldraw[fill=lightyellow, rounded corners = 2pt] (4,5) rectangle (4.9,5.9);
\filldraw[fill=lightyellow, rounded corners = 2pt] (5,5) rectangle (5.9,5.9);
\filldraw[fill=lightgreen, rounded corners = 2pt] (6,5) rectangle (6.9,5.9);

 \node[text width=2.5cm,anchor = east] at (-0.02,6.45)  [font=\small]
    {  optimized trade execution};
\filldraw[fill=lightred, rounded corners = 2pt] (0,6) rectangle (0.9,6.9);
\filldraw[fill=lightyellow, rounded corners = 2pt] (1,6) rectangle (1.9,6.9);
\filldraw[fill=lightgreen, rounded corners = 2pt] (2,6) rectangle (2.9,6.9);
\filldraw[fill=lightgreen, rounded corners = 2pt] (3,6) rectangle (3.9,6.9);
\filldraw[fill=lightyellow, rounded corners = 2pt] (4,6) rectangle (4.9,6.9);
\filldraw[fill=lightgreen, rounded corners = 2pt] (5,6) rectangle (5.9,6.9);
\filldraw[fill=lightyellow, rounded corners = 2pt] (6,6) rectangle (6.9,6.9);

\end{tikzpicture}\vspace{-8pt}
    \caption{Comparison of transaction reordering attacks mitigation approaches. Green indicates excellent performance in an area, yellow mediocre performance, and red unsatisfactory performance. Observe that while all approaches exhibit unsatisfactory performance in at least one measure, at least one approach excels in every measure.} \label{fig:summary}\vspace{0pt}
\end{figure}
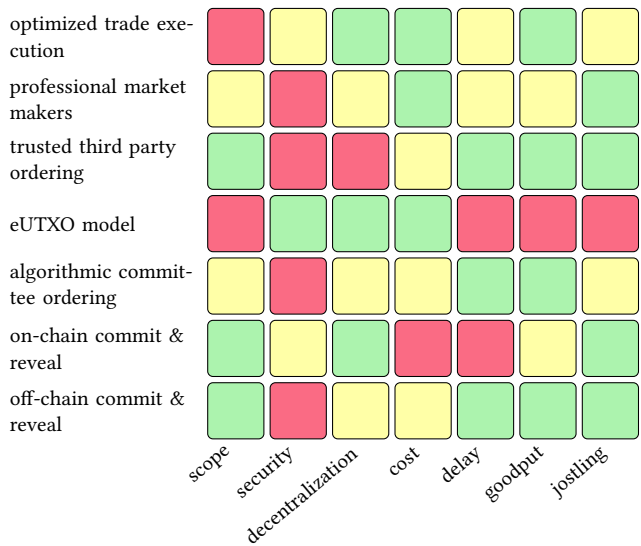

We conclude that, currently, no approach meets all the demands placed on a scheme to prevent transaction reordering manipulation attacks on a fully decentralized blockchain (cf. Figure~\ref{fig:summary}). More precisely, each approach displays poor performance in at least one of our seven measures. Instead of solely analyzing a protocol's overall or average performance across our measures, we also want to focus on the distinct impacts of an approach demonstrating poor performance in a particular measure in the following. While we consider all measures essential to judge a transaction reordering prevention scheme, we do not necessarily attribute them the same importance. 

Approaches that solely display poor performance in terms of scope, such as optimized trade execution (cf. Section~\ref{sec:optimizedtradeexecution}), might act well as temporary fixes. Optimized trade execution displays satisfactory performance in all other categories (cf. Figure~\ref{fig:summary}) and is therefore, well-equipped to function as a temporary fix -- at least in those areas that individual schemes tackle. It is, however, apparent that optimized trade execution will not cease transaction reordering manipulations, and new attack possibilities not addressed by the schemes are likely to arise over time. With its limited scope, it cannot tackle the broad and ever-changing landscape of BEV on the blockchain and can only be viewed as a temporary fix. 

On the other hand, approaches with unsatisfactory goodput or delay exhibit high and unacceptable performance losses for DeFi applications. Only evident by the original design of SundaeSwap, utilizing the eUTXO model (cf. Section~\ref{sec:eutxo}), never being implemented. Such an implementation would have led to both inadmissible delays and goodput levels. Similarly, we expect schemes with satisfactory goodput levels but unsatisfactory delays will be ill-equipped to meet the demands placed on DeFi. Therefore, we don't expect a widespread adaptation of on-chain committee ordering (cf. Section~\ref{sec:onchainreveal}). If DeFi wants to become a real alternative to its centralized counterpart, it simply cannot accept these performance losses.

The remaining four approaches all exhibit unsatisfactory performance in terms of security (cf. Figure~\ref{fig:summary}). Together with decentralization, poor performance in this measure is most at odds with the fundamental design principles of the blockchain. We further note an approach's security can never exceed its level of decentralization. Attacks from those in charge of ordering become easier as decentralization decreases and thereby decreases security. While algorithmic committee ordering (cf. Section~\ref{sec:committeeordering}), off-chain commit \& reveal (cf. Section~\ref{sec:offchainreveal}) and the utilization of professional market makers (cf. Section~\ref{sec:redesignDEX}), maintain at least some decentralization, this does not translate to increased security. A byzantine committee can easily perform transaction reordering manipulation to extract BEV in the two approaches that rely on a committee: algorithmic committee ordering and off-chain commit \& reveal. Utilizing professional market makers (cf. Section~\ref{sec:redesignDEX}), on the other hand, only excels in terms of cost and jostling. Otherwise, it is subordinate to off-chain commit \& reveal. We further note that while off-chain commit \& reveal's benefit is that the transaction contents only become visible upon ordering, which we expect to lead to less jostling, the protocol does not hinder a byzantine committee from decoding the message before ordering them. We currently see the greatest promise in off-chain commit \& reveal to tackle BEV with a broad scope, but have concerns regarding its security. 

Finally, the scheme that displays the highest level of centralization and therefore does not fare well in terms of security, trusted third party ordering (cf. Section~\ref{sec:privateordering}), is seeing the highest levels of adoption. Trusted third party ordering not only tackles all reordering attacks but also borrows excellent performance benefits in terms of scope, goodput, delay, and jostling from traditional exchanges. However, this comes at the expense of security and decentralization. We also want to mention that trusted third party approaches, if not utilized by all transactions, can be used and are currently used to perform transaction reordering attacks~\cite{2022flashbotsmev}. Still, we observe it being implemented at the fastest pace. 

Given this development, the question of whether centralized ordering has a place in DeFi should be raised sooner rather than later. Not only does it conflict with the idea of a fully decentralized blockchain, but the blockchain also does not protect users in case a trusted third party misbehaves. In traditional finance, transactions are also ordered by a trusted third party, typically an exchange, for example. However, users have the added regulatory protection restricting the exchange from manipulating the ordering in traditional finance.

Even though none of the current approaches to tackle transaction reordering manipulations can fulfill all the demands placed on them by the blockchain's ecosystem, it is very promising that, at the same time, multiple approaches that excel in any of our measures exist. We are, therefore, hopeful that a suitable approach will emerge in the future. Such a novel approach to prevent transaction reordering manipulations, thus, can and should leverage these existing ideas and learn from them in the areas where they excel.

\section{Conclusion}

The successful and efficient mitigation of transaction reordering manipulations on blockchains remains a challenge. While many approaches to tackle the problem are surfacing, there is currently no approach that can meet all requirements a fully decentralized blockchain places on a mitigation scheme. Thus, we hope for the search for a scheme preventing transaction reordering manipulations on the blockchain to continue. Especially, as we currently observe a widespread adoption of trusted third party orderings that conflict with the most fundamental idea of a fully decentralized blockchain, we hope that a suitable alternative to protect users against BEV will emerge. The direction of this research has the potential forever impact the future of blockchains.

Further, an approach that successfully prevents transaction reordering attacks on DeFi without impacting the remaining ecosystem negatively would outperform the current state of front-running mitigation in traditional finance. Not only would it remove all trust requirements, but preventing front-running practices from occurring is better than hoping that such practices are identified and prosecuted once they occurred.
\bibliographystyle{ACM-Reference-Format}
\bibliography{reference}
\balance



\end{document}